# Real-Time Magnetic Field Sensing based on Microwave Frequency Modulated Photocurrent of Nitrogen-Vacancy Centers in Diamond

Xuan-Ming Shen[1†], Qilong Wu[1†], Huihui Yu[1†], Pei-Nan Ni[1], Qing Lou[1], Chao-Nan Lin[2], Xun Yang[1*], Chong-Xin Shan[1*], Yuan Zhang[1,2*]

[1] Henan Key Laboratory of Diamond Materials and Devices, Key Laboratory of Material Physics, Ministry of Education, School of Physics, Zhengzhou University, Zhengzhou 450052, China
[2] Institute of Quantum Materials and Physics, Henan Academy of Sciences, Zhengzhou 450046, China

While photoelectric detection of magnetic resonance (PDMR) can be applied to miniaturize nitrogen-vacancy (NV) center-based quantum sensors, real demonstration of PDMR-based magnetic field sensing remains as a distinctive challenge. To tackle this challenge, in this article, we fabricate diamond samples with electrodes and microwave antenna on the surface, and realize PDMR by detecting photocurrent in picoampere range via various lock-in amplifying modes. We obtain a theoretical and experimental sensitivity 397 nT/$\sqrt{\text{Hz}}$ and 921 nT/$\sqrt{\text{Hz}}$ of magnetic field detection in DC~10 Hz range with a laser intensity and microwave frequency modulation mode, respectively, and demonstrate for the first time, a real-time tracking of alternating magnetic field with a standard deviation of 1.5 µT. Furthermore, we investigate systematically the dependence of the PDMR contrast, linewidth and the sensitivity on the laser and microwave power, and find a perfect agreement with a master equation-based theory. Thus, our results represent a critical step forward in transitioning PDMR from a spectroscopic technique to a practical sensing modality.

## I. INTRODUCTION

Nitrogen-vacancy (NV) center, as a prototype solid-state qubit, has attracted considerable attention in the fields of quantum information [1,2], computation [3-5] and sensing [6-8], owing to its long spin coherence time at room temperature [9,10], convenient spin initialization and readout with optical means. Beyond the conventional readout via spin-dependent photoluminescence, other readout schemes have also been actively investigated in recent years, such as the readout via photocurrent [11], infrared absorption [12], and microwave transmission [13]. Among these schemes, electrical readout as explored in electrical/photocurrent detection of magnetic resonance (EDMR/PDMR) technique attracts particular interest [11,14,15], as it removes the requirement of optics and electronics to collect the NV center fluorescence, and to convert it to electric signal [16-19], which thereby can facilitate the integration and miniaturization of the NV center-based quantum sensors [20].

The development of PDMR has followed the path of the photon readout scheme, achieving the milestones of detecting NV spin ensembles [15,21,22], single NV spins [23,24] and nuclear spins [25,26]. All these are enabled by a better understanding of the laser-induced charge switch of negative and neutral NV centers [27]. While PDMR-based magnetic detection has been reported before [20,28,29], these studies employ mainly lock-in amplifier (LIA) with either laser or microwave modulation to detect the weak photocurrent and to obtain the PDMR spectra, and estimate *theoretically* the magnetic field sensitivity through theoretical formulas. Although PDMR was also demonstrated recently with other color centers [30-32], the corresponding magnetic field sensing still lacks a systematic characterization and thorough validation. Thus, to further advance PDMR technology and its practical applications, it is essential to develop the LIA technique based on microwave frequency modulation to achieve a dispersive PDMR spectrum, to validate the theoretically calculated magnetic field sensitivity with amplitude spectral density measurement, and to demonstrate real-time magnetic field sensing under practical conditions.

To address the above challenges, in the present article, we carry out *an experiment-theorical joint study* on the PDMR based magnetic field sensing by fabricating a diamond sample with integrated electrodes and microwave antennas on the surface, and detecting weak photocurrent in

cxshan@zzu.edu.cn
yangxun9103@zzu.edu.cn
* yzhuaudipc@zzu.edu.cn

picoampere range via various lock-in amplifying modes [33]. We obtain a *theoretically estimated and experimentally validated* sensitivity of 397 nT/$\sqrt{Hz}$ and 921 nT/$\sqrt{Hz}$ for magnetic field detection with a laser intensity and microwave frequency modulation mode, respectively. We demonstrate for the first time, as far as we know, a real-time tracking of alternative magnetic field with a standard deviation of 1.5 µT. Importantly, we carry out a systematic study of the dependence of the PDMR contrast, linewidth and the sensitivity on the laser and microwave power, and achieve a perfect agreement with a master equation-based theory. Based on this agreement, we further estimate that the magnetic field sensitivity might be improved to tens nT/$\sqrt{Hz}$ by using a diamond sample featuring higher NV concentration. Thus, our results represent a critical step forward in transitioning PDMR from a spectroscopic technique to a practical sensing modality, and the agreement between theory and experiment provides a solid foundation for further optimizing PDMR-based sensors.

## II. NV CENTER AND EXPERIMENTAL SETUP

Figure 1(a) shows the simplified schematic of the setup and the sample, which are further detailed in Appendix A and B. The sample is a MPCVD diamond grown in our lab, and features two pairs of gold interdigitated electrodes and a gold microbelt as a microwave antenna on the surface. A 532 nm laser is focused between the gold electrodes through an objective, and the NV fluorescence is detected in an inverted direction. The microwave is first amplified, and then delivered through the antenna near the electrodes. A DC voltage is applied between the electrodes, and the photocurrent is detected with a lock-in amplifier.

Before presenting our results, we recapture shortly the principle of PDMR-based magnetic field sensing. NV center is a point defect formed by a substitutional nitrogen atom and an adjacent vacancy in the diamond crystal [Fig. 1(b)]. NV center can be either in a neutral form ($NV^0$) or a negatively charged form ($NV^-$), where $NV^-$ has a triplet ground and excited state $^3A_2$, $^3E$ and two singlet excited states $^1A_1$ and $^1E$, and $NV^0$ has a doublet ground state $^2E$ and excited state $^2A_1$ [Fig. 1(c)]. In addition, the fine structure of triplet states can be labeled by magnetic projection numbers $m_s$ = 0, ±1, and the degenerate $m_s$ = ±1 states are about 2.87 GHz and 1.42 GHz above the $m_s$ = 0 state for the $^3A_2$ and $^3E$ state.

Under 532 nm laser illumination, the $NV^-$ can be excited to the $^3E$ state and then be converted to $NV^0$ by absorbing one photon and two photons, leaving a free electron in the conduction band of diamond. Similarly, $NV^0$ can be excited to the $^2A_1$ state and then be converted to $NV^-$ by absorbing one and two photons, generating a free hole in the valence band of diamond. The excited $NV^-$ and $NV^0$ can decay to the ground state through spontaneous emission, while the excited $NV^-$ can also decay through non-radiative processes involving triplet-singlet intersystem crossing (ISC) and the decay from the $^1A_1$ state to the $^1E$ state. Because the ISC rate from the $^3A_2$, $m_s$ = ±1 states is faster than that from the $^3A_2$, $m_s$ = 0 state in the $NV^-$ center, the non-radiative process leads to large population on the $m_s = 0$ state and a spin-dependent photocurrent and fluorescence. When a microwave at around 2.87 GHz is applied to the NV centers, the population is transferred from the $^3A_2$, $m_s = 0$ level to the $^3A_2$, $m_s = \pm1$ level, and the corresponding photocurrent and fluorescence will be reduced, forming the principles behind PDMR and Optical Detection of Magnetic Resonance (ODMR). In the presence of a weak magnetic field, the $m_s = +1$ and $m_s = -1$ levels can be shifted slightly upwards and downwards in energy, respectively, and the PDMR and ODMR spectra become modified, from which the magnetic field can be inferred.

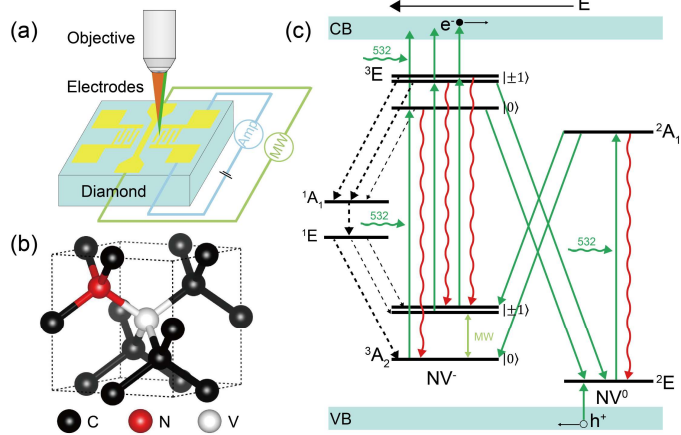

FIG. 1. Setup and operating principle. (a) Simplified diagram of the setup, where a 532 nm laser is focused between two electrodes and near a microwave antenna on a MPCVD diamond sample, and the photocurrent is detected through a lock-in amplifier. (b) Atomic structure of NV centers, where black, red and white spheres represent carbon atoms, nitrogen atom and vacancy, respectively. (c) Energy level diagram and rich processes of negatively charged ($NV^-$) and neutral NV centers ($NV^0$), where laser excitation (green arrows) leads to the fluorescence (red arrows) and the generation of electrons in conduction band (CB) and holes in valence band (VB) through charge-state conversion. For more information see the text.

In many studies on PDMR and ODMR, lock-in amplifier is often adopted to extract the signal from the noisy photocurrent or fluorescence, where the laser intensity [15, 28], the microwave intensity and frequency [17, 34-36] or even the external magnetic field can be modulated. Among these modes, microwave frequency-modulation is often adopted to implement the ODMR-based real-time detection of alternating magnetic field, since the resulting dispersive spectrum shows a linear scaling of the lock-in output to the microwave carrier frequency under resonant condition, which allows the direct conversion of the lock-in output to the measured magnetic field [32,37]. However, as far as we know, such a technique has not been explored in the studies of PDMR, and thus there is no report so far on the PDMR based real-time magnetic field detection.

### III. EXPERIMENTAL AND THEORETICA RESULTS

In this section, the photocurrent-voltage and photocurrent-laser power measurement on the diamond sample were characterized firstly. Then, the lock-in PDMR with the laser intensity modulation was presented, and the sensitivity of the magnetic field detection was estimated theoretically. After that, the lock-in PDMR with the microwave frequency modulation was discussed, and the theoretical sensitivity was verified through the amplitude spectral density measurement. The real-time tracking of alternative magnetic field was also demonstrated. Finally, the dependence of the system performance on the laser and microwave power was studied, and the experimental results were compared with a master equation-based theory.

#### A. Photocurrent-Voltage and -Laser Power Characteristics

In the experiment, we firstly characterize the fabricated device with the photocurrent-voltage (I-V) and the photocurrent-laser power (I-P) curves [Figs. 2(a) and 2(b)]. To this end, we apply a DC voltage and a continuous laser illumination on the device, and setup the low-noise current amplifier with a low-pass filter cut-off frequency of 0.03 Hz and a proper transimpedance gain of 1 nA/V, and then average the DC photocurrent with the oscilloscope integrated in the LIA for 5 s.

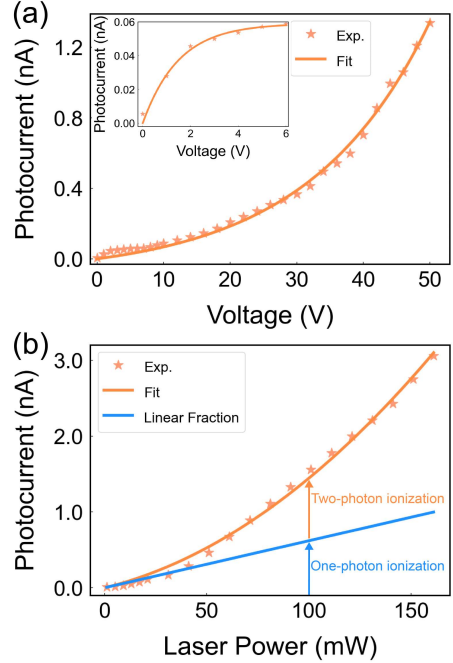

FIG. 2. Electric characterization of the NV center diamond sample. (a) Photocurrent as function of the applied DC voltage for given laser power 100 mW (the inset presents the magnified results in the voltage range of 0–6 V). (b) Photocurrent as function of the laser power for an applied voltage of 50 V. The experimental data (orange star) are fitted with nonlinear functions stated in the main text (orange curves). In panel (b), the blue curve is the linear fraction of the photocurrent.

Figure 2(a) illustrates the I-V characteristics for 100 mW laser illumination, which shows a raising and saturated trend in the low voltage range (inset) and an exponentially increased behavior in the high voltage range. Inspired by our previous study [38] and a recent study [39], we realize that the electrode-diamond-electrode structure in our device forms two back-to-back connected Schottky junctions. At low voltage, the current is determined by the backward biased junction, and follows the expression $J_1 = A_1(1 - e^{-U/U_1})$. At high voltage, the depletion region of this junction increases and touches that of another junction.

As a result, the two junctions merge as single Schottky junction, and the current is determined by the forward biased expression $J_2 = A_2(e^{U/U_2} - 1)$. Using the two expressions to fit the I-V curve in the low- and high voltage range, we obtain the amplitudes $A_1 = 59.1 \pm 2.3$ pA, $A_2 = 93.31 \pm 7.1$ pA, and the saturated and threshold voltage $U_1 = 1.5$ V, $U_2 = 18.28$ V. Since the large photocurrent under high bias can be more easily detected with the LIA, a DC bias voltage of 50 V is used for the subsequent measurements.

Figure 2(b) shows that the photocurrent $J$ increases firstly slowly with the laser power $P$ for weak laser illumination, but increases much faster for the laser power exceeding 50 mW. Since the optically induced carriers from the NV centers originate from two-photons ionization of NV$^-$ and two-photons recombination of NV$^0$, the photocurrent is expected to scale quadratically with the laser power. However, due to the presence of other photon-active defects in the diamond, the photocurrent contains also a background contribution, which should scale linearly with the laser power. Thus, we fit the data in Fig. 2(b) with the expression $J = aP + bP^2$ to obtain the fitting parameters $a = 6.602$ pA/mW and $b = 0.079$ pA/mW$^2$. The laser power, where the linear term equals to the quadratic term, can be estimated as $\frac{a}{b} \approx 86$ mW.

Thus, in the following study, we utilize the laser power of 100 mW such that the photocurrent from the NV centers should dominate.

**B. Laser Intensity-modulated Lock-in PDMR**

In the experiment shown in Fig. 3, the 100 mW laser beam was modulated by a 321 Hz square wave, and a DC voltage of 50 V was applied between the electrodes. After amplifying the photocurrent by the pre-amplifier with a 0.3 kHz ~1 kHz band-pass filter and a transimpedance gain of 1 nA/V, the output voltage was demodulated with LIA. By integrating the photocurrent of 4 s for the microwave near and far-detuned from 2.87 GHz with a power of 26 dBm, and then calculating their contrast, the PDMR spectrum was obtained, see blue stars in Fig. 3(a). By fitting this curve with a Lorentzian function

$$L(f) = C \frac{(\Gamma/2)^2}{(f - f_0)^2 + (\Gamma/2)^2} + C_0, \quad (1)$$

we extract a maximal contrast $C = 12.1\%$, a center frequency $f_0 = 2869$ MHz, and a linewidth $\Gamma = 19.5$ MHz. Then, the shot-noise limited sensitivity can be estimated according to the following expression

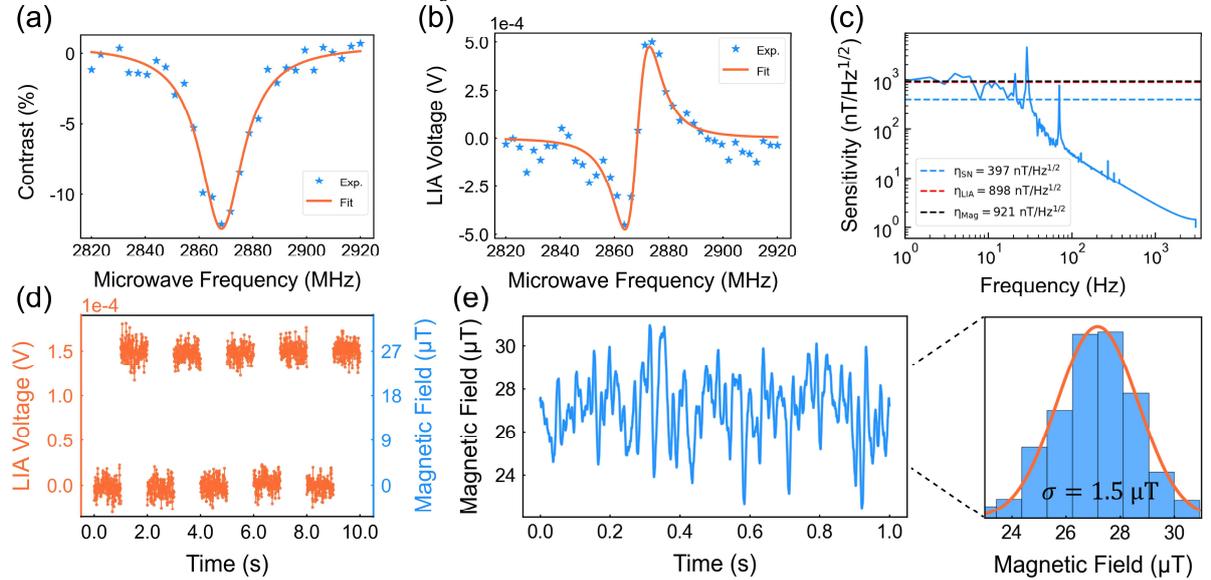

FIG. 3. PDMR and magnetic field sensing. (a) PDMR spectrum with the laser intensity modulation. (b) Lock-in voltage as function of the microwave frequency with microwave frequency-modulation. Blue stars are the experimental results, and the orange curves are the fits according to the expressions stated in the text. (c) Sensitivity of the magnetic field detection based on frequency-modulated PDMR as function of the frequency (blue noisy curve), and the corresponding theoretical estimations (dashed vertical lines). (d) Real-time tracking of an alternating magnetic field (right axis), as derived from the lock-in voltage (left axis). (e) Enlarged trace of the magnetic field signal from the final 1-second segment of (d) and the histogram of the steady-state value with a standard deviation of 1.5 µT. For the panels (a-e), the DC voltage is 50 V, the laser power is 100 mW, the microwave power is 26 dBm, and the frequency-modulation depth is 1 MHz. For more details, see the text.

$$\eta_{SN} = \frac{4h}{3\sqrt{3}g\mu_B}\frac{\Gamma}{C\sqrt{R}}. \quad (2)$$

Here, $h$ is Planck constant, $g \approx 2.003$ is the electron $g$ factor, $\mu_B$ is the Bohr magneton, and $R$ is the rate of photocarriers detection. Note that here $R$ corresponds to the photocurrent detected with the LIA when the laser or microwave are modulated [Fig. 9(e)], and it should not be confused with the photocurrent measured without modulation, as shown in Fig. 2(b). Using the fitted parameters, a magnetic field sensitivity of $\eta_{SN} = 397$ nT/$\sqrt{\text{Hz}}$ can be estimated, as indicated by the blue dashed line in Fig. 3(c).

### C. Microwave Frequency-modulated Lock-in PDMR

Since the above PDMR spectrum with a dip does not show a linear dependence on the microwave frequency, it is not convenient to utilize it for weak magnetic field sensing. To solve this problem, we explore the lock-in detection based on the microwave frequency-modulation, and obtain a PDMR spectrum with a dispersive shape as shown in Fig. 3(b), where the lock-in voltage shows a linear scaling with the microwave around 2870 MHz. By fitting the curve with the following expression [41]

$$L(f) = -\frac{32}{3\sqrt{3}}C\frac{(f-f_0)\Gamma^3}{[4(f-f_0)^2 + \Gamma^2]^2}, \quad (3)$$

we estimate a maximal amplitude $C = 5 \times 10^{-4}$ V, a center frequency $f_0 = 2868.43$ MHz, a linewidth $\Gamma = 15.257$ MHz. For the linear region near the resonance, we estimate $L \approx S(f - f_0) = S\gamma_e B$ with a slope of $S = \frac{32\sqrt{3}}{9}\frac{C}{\Gamma} = 2.018 \times 10^{-4}$ V/MHz. Here, $\gamma_e = g\mu_B = 2.8$ MHz/Gauss is the gyromagnetic ratio of the electron. The inversed expression $B = L/(S\gamma_e)$ can be explored to infer the weak magnetic field in a real-time fashion. Then, the sensitivity of the magnetic field detection can be theoretically estimated through the expression [42]

$$\eta_{LIA} = \frac{\sigma}{\gamma_e S\sqrt{2f_{ENBW}}} \quad (4)$$

as 898 nT/$\sqrt{\text{Hz}}$ [red dashed line in Fig. 3(c)] with the standard deviation of the LIA signal $\sigma = 2.274 \times 10^{-5}$ V and the equivalent noise bandwidth of the LIA $f_{ENBW} = 10$ Hz.

To verify experimentally the theoretically estimated sensitivity, the lock-in voltage $\{V_i\}$ was recorded under the resonance condition for 5 s, and was then translated to the magnetic field $\{B_i\}$. Then, the data was chopped into segments of $\tau = 1$ s long, and discrete Fourier transform $\mathcal{F}$ was performed to calculate the amplitude spectral density $\frac{|\mathcal{F}|}{S\gamma_e}\sqrt{\tau}$ [33,34,43-45]. The obtained spectrum is shown as the blue noisy curve in Fig. 3(c), and indicates a sensitivity of $\eta_{Mag} \approx 921$ nT/$\sqrt{\text{Hz}}$ (black dashed line) in the frequency range 1-10 Hz. Such a sensitivity is close to that estimated with Eq. (4), but is about 2 times smaller than that estimated with Eq. (2), which can be attributed to the electronic noise of the measurement setup.

By exploring the linear region in the microwave frequency-modulated PDMR, we can now demonstrate the real-time tracking of the magnetic field [Fig. 3(d)]. To this end, we place an electromagnet beside our setup, and modulate the applied magnetic field by changing the electromagnet current periodically with a frequency 1 Hz. An alternation of the lock-in voltage between 0.15 mV and 0 mV was observed, which corresponds to a magnetic field variation between 27 μT and 0 μT. To further evaluate the signal stability, the data for the last 1 second were magnified, and a histogram with a mean 27 μT and a standard deviation of about 1.5 μT was extracted [Fig. 3(e)], indicating the practical magnetic field resolution of the current system.

### D. Microwave Frequency-modulated Lock-in PDMR

After demonstrating the PDMR-based magnetic field sensing, we examine now with Fig. 4 the dependence on the laser power (a-c) and the microwave power (d-f). Here, we set the lowest laser and microwave power as 70 mW and 20 dBm so that the PDMR spectrum can be faithfully determined, and the orange stars are the experimental data. As the laser power $P$ increases from 70 mW to 130 mW, the contrast $C$ increases firstly from 11.1%, approaches the maximum 12.1% for $P = 100$ mW, and then decreases to 10.3% [Fig. 4(a)]. In contrast, the linewidth $\Gamma$ increases mountainously from about 14.1 MHz to 21.9 MHz [Fig. 4(b)]. The behavior of the contrast can be understood through the competition of the optically induced spin polarization and the microwave absorption-induced population transfer, and the increased linewidth is caused by the increased rate of the former process [11,46,47]. Interestingly, the sensitivity $\eta$ decreases fast from 600 nT/$\sqrt{\text{Hz}}$ to about 400 nT/$\sqrt{\text{Hz}}$ as the laser power increases from 70 mW to 100 mW, and then becomes saturated [Fig. 4(c)]. Based on Eq. (2), the flat feature can be understood by the cancelation of the quadratically increased carrier generation rate $R$ [Fig. 9(e)] to the increased ratio of the linewidth and contrast. In strong contrast, as the microwave power increases from 20 dBm to 26 dBm, the contrast increases from 6% to 12% [Fig. 4(d)], and the linewidth increases from 14 MHz to 20 MHz [Fig. 4(e)], which both can be attributed to the increased stimulated microwave absorption. As a result, the sensitivity shows a monotonous decay from 550 nT/$\sqrt{\text{Hz}}$ to 400 nT/$\sqrt{\text{Hz}}$ for increasing microwave power [Fig. 4(f)].

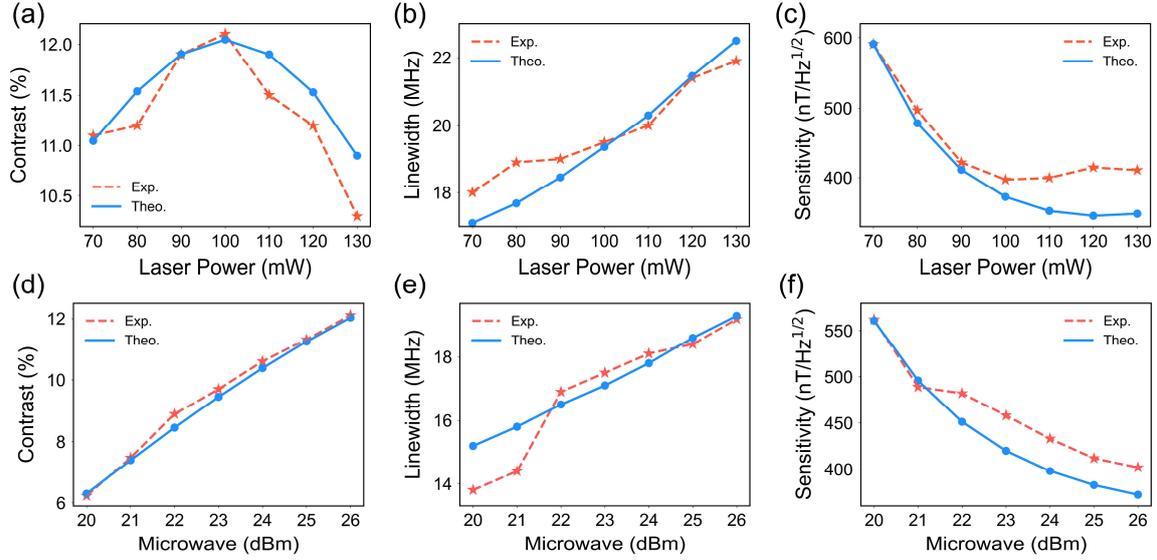

FIG. 4. Dependence of the magnetic field sensing on the laser (a-c) and microwave power (d-f). The results are the PDMR contrast (a,d), linewidth (b,e) and the sensitivity estimated with Eq. (2) (c,f). Orange stars are the experimental data, and the blue dots are the results based on the master equation-based theory. In the panels (a-c), the microwave power is 26 dBm. In the panels (d-f), the laser power is 100 mW.

To understand further the experimental results, theoretical studies were carried out for the current system by adopting a quantum master equation to describe all the processes of NV centers outlined in Fig. 1(b), and using parameters reported in the previous literature and characterized with a commercial setup [48-50], see Appendix D for more details. Furthermore, we have accounted for the number of NV centers $2.551 \times 10^3$, the conversion from the laser power to the optical pumping rate, and the conversion from the microwave power to the microwave magnetic field felt by the NV centers. The first parameter is estimated with the size of the laser focus and the NV concentration, and the second one is estimated from the laser power dependence of the fluorescence. Since the last parameter depends on the microwave field distribution and the relative position of the NV centers to the antenna, it is rather difficult to estimate and thus is considered as a fitting parameter. In the theoretical studies, we simulate the PDMR spectra to extract the contrast and linewidth, and then estimate the sensitivity according to Eq. (2). The theoretical results are presented by the blue dots in Fig. 4, and show a remarkable agreement with the experimental results.

Because of the good agreement with the experiment, the proposed theory can be further explored to optimize the PDMR device to achieve a better sensitivity. For example, by using an electron-irradiated diamond sample, the NV concentration can be increased by three orders of magnitude to few ppm. Then, the same orders of magnitude increase of the carrier's detection rate might lead to an improvement of sensitivity by 32 times to tens nT/$\sqrt{\text{Hz}}$ according to Eq. (2).

## V. CONCLUSIONS

In summary, we have carried out an experiment-theory joint study on the PDMR-based magnetic field sensing. By fabricating a diamond sample with electrodes and microwave antenna on the surface and using a home-built setup, we obtain the PDMR spectrum by detecting photocurrent in picoampere range with various lock-in amplifying modes. We have obtained a theoretically shot-noise limited sensitivity of 397 nT/$\sqrt{\text{Hz}}$ with the laser power modulation mode, and an experimentally electronic noise-limited sensitivity 921 nT/$\sqrt{\text{Hz}}$ with the microwave frequency-modulation mode. Importantly, for the first time, as far as we know, we demonstrate a real-time tracking of alternating magnetic field with a standard deviation of 1.5 μT. To optimize the sensitivity further, we investigate systematically the dependence of the PDMR contrast, linewidth and the sensitivity on the laser and microwave power, and achieve a perfect agreement with a master equation-based theory. Based on the theory, we predict that the sensitivity can be improved to tens nT/$\sqrt{\text{Hz}}$ with an electron-irradiated diamond sample of higher NV concentration. Thus, our results represent a critical step forward in transitioning PDMR from a spectroscopic technique to a practical sensing modality, and the agreement between theory and experiment provides a solid foundation for further optimizing PDMR-based sensors.

## ACKNOWLEDGMENTS

This work was supported by the National Key R&D Program of China Grant No. 2024YFE0105200, the Scientific Research Innovation Capability Support Project for Young Faculty Grant No. SRICSPYF-BS2025008, the National Natural Science Foundation of China Grant Nos. 62475242, 12422413, and U25A20192, the Joint Fund of Henan Province Science and Technology R&D Program Grant No. 245200810005, the Cross disciplinary Innovative Research Group Project of Henan Province Grant No. 232300421004. Xuan-Ming Shen carried out the results under the supervision of Yuan Zhang. All authors contributed to the analysis and writing of the manuscript.

**Conflict of interest** The authors declare that they have no conflict of interest.

## APPENDIX A: DETAILS OF EXPERIMENTAL SETUP

In this appendix, we provide further details on our experimental setup, which consists of an optical and electronic module in (Fig. 5). The optical module is constructed with an optical cage system [Fig. 5(a)], and includes a solid-state laser (532 nm wavelength, 300 mW power, Changchun New Industries, MGL-S-532), two mirrors (532 nm, JC OME1-1-R2), a dichroic mirror (400-633 nm reflection, 685-1600 nm transmission, JC OFD1LP-650), an objective lens (50X, 0.55 NA, Shenzhen Beiyinhu Optics Co., Ltd., 50X SLMplan), a long-pass filter (650 nm, JC OFE1LP-650), a plano-convex lens (50 mm focal length, JC OLC220136), a beam splitter (400-700nm, OSB2555-T2), a CMOS industrial camera (MER2-160-249U3M-HS-6P, Daheng), and two three-axis translation stages. The entire optical path is mounted on a $40 \times 70$ cm$^2$ breadboard, facilitating the portability and installation of the integrated optical system. The 532 nm laser beam is focused through the objective on the diamond sample after being first reflected by two mirrors and a dichroic mirror. The NV fluorescence is collected in the backward direction through the objective lens, then is passed sequentially through a dichroic mirror and a long-pass filter for filtering. Subsequently, the beam was split by a beam splitter into two beams, which are focused separately by two lens onto the camera for imaging and a photodetector (PD). One translation stage is connected to the diamond sample, while another translation stage is equipped with a square magnet.

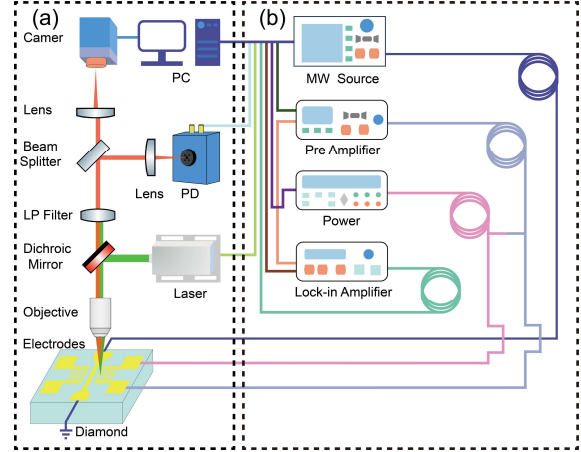

FIG. 5. Detailed schematics of the optical (a) and electronic part (b) of the experimental setup.

The electronic module as shown in Fig. 5(b) includes a microwave source (DSG836A, Rigol), a power amplifier (KDT2038PA-045, Kedite), a lock-in amplifier (LIA001M, Guoyi Quantum), a low-noise current amplifier (SR570, Stanford Research Systems), a programmable power supply (IT6322C, ITECH), and a laser driver (532 nm, Changchun New Industries). The microwave source can output continuous-wave, amplitude-modulated, frequency-modulated microwave. The microwave is amplified by the power amplifier, and is then delivered to the sample via a coaxial cable. The laser driver outputs either a constant current for continuous-wave laser operation or a frequency-modulation current for modulated laser output. The programmable power supply, with its two channels connected in series, provides an output voltage up to 60 V, delivers power to the sample through cables. The laser driver, the microwave source, the LIA, and the programmable power supply are all connected to a computer with USB cables, and controlled by programs based on Python language.

## APPENDIX B: PREPARATION AND INTEGRATION OF DIAMOND SAMPLE

In this appendix, we explain the sample preparation and its integration with the experimental setup (Fig. 6). The diamond sample is grown in our laboratory via a MPCVD method, and is characterized with a Quantum Diamond Spin Spectrometer (Chinainstru & Quantumtech Co., Ltd, Diamond-II Studio), see Fig. 7.

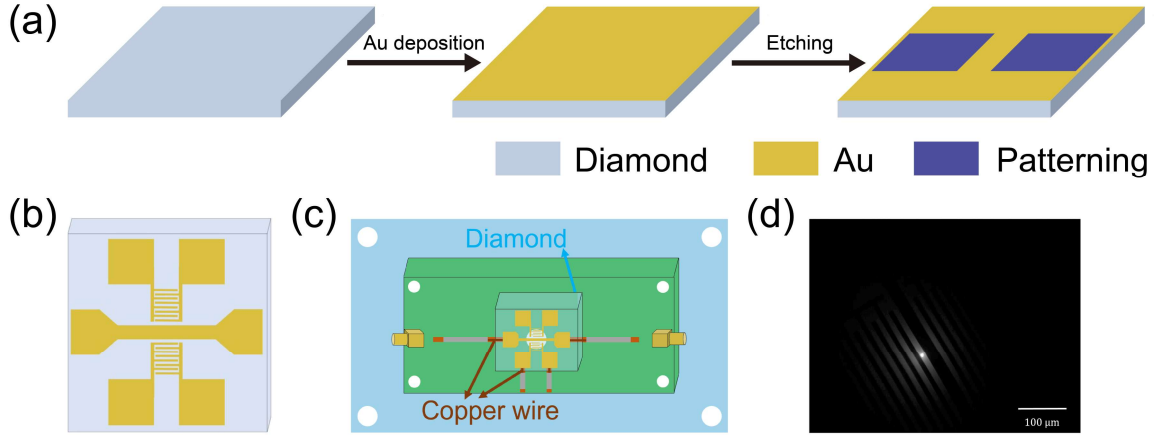

FIG. 6. Preparation of diamond sample. (a) Schematic diagram of the device fabrication process. (b) Layout of the fabricated sample. (c) Integration of the sample in the sample hold. (d) Fluorescence image showing the position of laser focus between the electrodes and near to the microwave antenna.

The procedure to prepare the sample is shown in Fig. 6(a). First, the sample is immersed in a mixture of concentrated hydrochloric acid (HCl) and concentrated nitric acid ($HNO_3$) at a volume ratio of 3:1 for 8 hours, and is then further placed in an acidic mixture of sulfuric acid ($H_2SO_4$) and potassium nitrate ($KNO_3$) at 250°C for 30 minutes for cleaning and oxidation. The sample is thoroughly rinsed with ultrapure water to remove residual acids, and is then ultrasonically cleaned in acetone, ethanol, and deionized water for 30 minutes to remove surface impurities. Second, a 100 nm gold film is deposited onto the diamond surface via magnetron sputtering, and two groups of interdigitated gold electrodes of 10 μm wide and 10 μm spacing and a gold belt of 25 μm wide are patterned onto the diamond surface using laser direct writing and photolithography equipment. Finally, the fabricated sample is annealed in a tube furnace at 510°C under vacuum for 3 hours to reduce the interface barrier between the diamond and electrodes.

Figure 6(b) presents a layout illustration of the fabricated diamond device, where six larger yellow pads indicate the patterned metal electrodes. The top two electrodes are connected to one pair of interdigital electrodes, while the bottom two electrodes are connected to another pair. The left and right electrodes are connected to a metal strip serving as a microwave antenna. The scale bar confirms that both the width and spacing of the interdigital electrodes are 10 μm. The width of the copper strip used as the microwave antenna is 25 μm, and its distance from the electrodes is 15 μm.

Figure 6(c) illustrates the integration of the sample with the sample holder. The backside of the sample was attached to the holder using double-sided adhesive tape. Subsequently, indium pellets were used to connect one end of two copper wires (with a diameter of 160 μm) to either the top or bottom pair of electrodes, while the other ends were soldered to the tin wires on the upper or lower side of the holder. Simultaneously, two additional copper wires were connected to the left and right electrodes using indium pellets, and these wires were soldered to the tin wires connected to the SMA interfaces on the left and right sides of the holder. Finally, the holder was assembled onto a 3-axis translation stage within the optical system, with the sample facing upward. The position of the sample relative to the objective lens was adjusted. The laser was then turned on, and fluorescence imaging was used to position the laser spot in the center of the electrodes and near the microwave antenna [Fig. 6(d)].

## APPENDIX C: EXTRA EXPERIMENTAL RESULTS

To determine several key parameters involved in the theoretical model, we have characterized our MPCVD diamond with a Quantum Diamond Spin Spectrometer (Chinainstru & Quantumtech Co., Ltd, Diamond-II Studio). We obtain firstly the confocal fluorescence image for a laser power 3 mW, and estimate the NV concentration as 3.3 ppb by comparing it with the photon count rate of single NV center [Fig. 7(a)]. Then, we apply an external magnetic ODMR spectrum for a microwave of 26 dBm power [Fig. 7(b)]. By concentrating on the leftmost dip, we further carry out Rabi oscillations experiment [Fig. 7(c)], and extract the $\pi/2$ and $\pi$ microwave time as 100.76 ns and 171.83 ns. Using these times, we carry out Ramsey, Hahn echoes and spin depolarization experiments, and extract the characteristic times $T_1 = 5$ ms, $T_2 = 15$ μs, $T_2^* = 901$ ns by fitting the curves with the proper expressions [Figs. 7(d)-7(f)]. Based the literature report on the relation of $T_2$ and nitrogen defect concentration [6], we estimate a nitrogen concentration of 1.5 ppm, leading to a NV⁻/N ratio of $2.2 \times 10^{-3}$. These parameters will be used in the latter theoretical model.

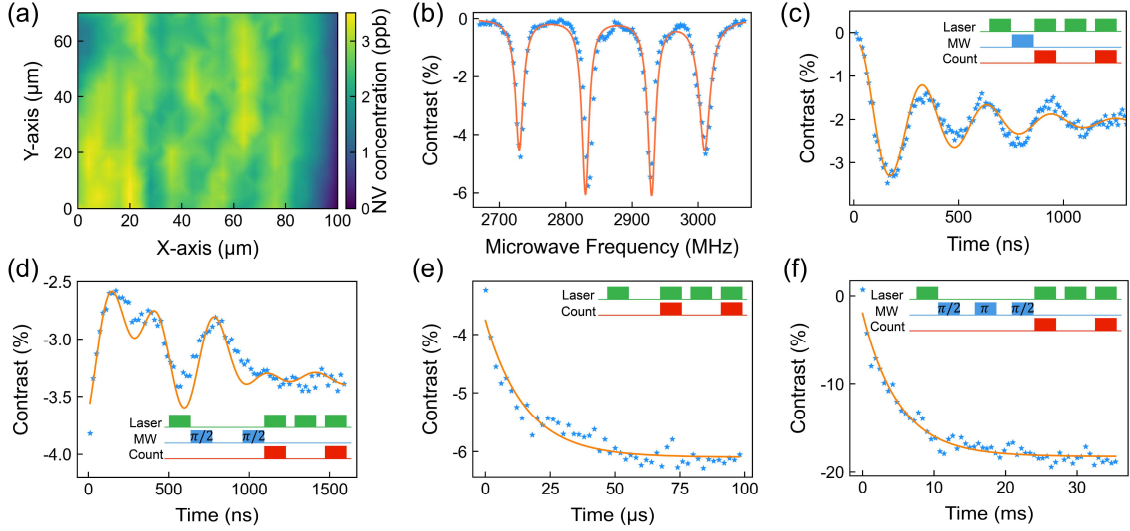

FIG. 7. Characterization of the MPCVD diamond sample. (a) NV concentration map of the fabricated diamond sample. (b) ODMR spectrum for an external magnetic field along one NV axis. (c,d,e,f) Results of the Rabi oscillations (c), Ramsey (d), Hahn echoes (e) and spin depolarization (f) experiment, respectively, with the pulse sequences of laser, microwave and readout shown in the insets. The blue dots are experimental results, and the orange solid curves are the fits.

## APPENDIX D: THEORY OF PDMR-BASED MAGNETIC FIELD SENSING

In this appendix, we present the theory for the PDMR-based magnetic field sensing, and the corresponding numerical results.

### 1. Quantum Master Equation

We describe the system with the reduced density operator $\hat{\rho}$, and establish the following master equation to describe the processes outlined in Fig. 1(c):

$$\frac{\partial}{\partial t}\hat{\rho} = -\frac{i}{\hbar}[\hat{H}, \hat{\rho}] - \mathcal{D}[\hat{\rho}]. \quad (D1)$$

The system Hamiltonian takes the form $\hat{H} = \sum_\alpha \hat{H}_\alpha = \sum_\alpha h\left[D_\alpha\left(\hat{S}_{z,\alpha}^2 - \frac{2}{3}\right) + \gamma_e \boldsymbol{B}\cdot\hat{\boldsymbol{S}}_\alpha\right]$ (with $\alpha = {}^3A_2, {}^3E$). The zero-filed splitting is $D_\alpha = 2.87$ GHz and 1.42 GHz for the triplet ground and excited level ${}^3A_2$, ${}^3E$, respectively. The spin vector $\hat{\boldsymbol{S}}_\alpha = \hat{S}_{\alpha,x}\boldsymbol{e}_x + \hat{S}_{\alpha,y}\boldsymbol{e}_y + \hat{S}_{\alpha,z}\boldsymbol{e}_z$ is defined with three components $\hat{S}_{\alpha,x}, \hat{S}_{\alpha,y}, \hat{S}_{\alpha,z}$ in the Cartesian coordinate system (with $\boldsymbol{e}_x, \boldsymbol{e}_y, \boldsymbol{e}_z$ as the unit vectors). $\gamma_e = 2.8$ MHz/Gauss is the electron gyromagnetic factor. The magnetic field $\boldsymbol{B} = B_x\boldsymbol{e}_x + B_y\boldsymbol{e}_y + B_z\boldsymbol{e}_z$ is specified by three components $B_x, B_y, B_z$, where the time-independent terms $B_x^{st}, B_y^{st}, B_z^{st}$ describe the static magnetic field, and the time-dependent terms $B_x^{mw}(t), B_y^{mw}(t), B_z^{mw}(t)$ are related to the microwave radiation. Within the space defined by the $m_s = +1, 0, -1$ spin levels of the triplet states ${}^3A_2, {}^3E$, the three spin components can be written as a matrix, and thus the Hamiltonian $\hat{H}_\alpha$ (with $\alpha = {}^3A_2, {}^3E$) can be also written as matrices

$$\frac{\hat{H}_\alpha}{h} = \begin{pmatrix} \frac{1}{3}D_{s,\alpha} + \gamma_e B_z & \frac{1}{\sqrt{2}}\gamma_e B_- & 0 \\ \frac{1}{\sqrt{2}}\gamma_e B_+ & -\frac{2}{3}D_{s,\alpha} & \frac{1}{\sqrt{2}}\gamma_e B_- \\ 0 & \frac{1}{\sqrt{2}}\gamma_e B_+ & \frac{1}{3}D_{s,\alpha} - \gamma_e B_z \end{pmatrix} \quad (D2)$$

with the abbreviations $B_\pm = B_x \pm iB_y$.

To solve Eq. (D1) numerically, we have associated the levels of the NV⁻ and NV⁰ center with an integer $l = 1,...,10$, see Fig. 8(a). For the NV⁻ center, the spin levels $m_s = 0, -1, +1$ of the ${}^3A_2$ and ${}^3E$ level are associated with the integers 1,2,3 and 4,5,6, while the levels ${}^1E, {}^1A_1$ level are labeled with 7,8. For the NV⁰ center, the ${}^2E, {}^2A_1$ level are associated with 9, 10. Then, we define the reduced density operator as a matrix within the space defined by these relabeled states $\{|l\rangle\}$. For a better notation, we introduce the projection operators $\hat{\sigma}_{l,l} = |l\rangle\langle l|$ and the transition operators $\hat{\sigma}_{l,l'} = |l\rangle\langle l'|$ ($l \neq l'$). In this case, we can rewrite the Hamiltonians as function of these operators (not shown), and rewrite the dissipation and dephasing processes explicitly as:

$$\mathcal{D}[\hat{\rho}] = -\sum_{l=1}^{3}(k_{l\to 3+l}\mathcal{D}[\hat{\sigma}_{3+l,l}]\hat{\rho} + k_{3+l\to l}\mathcal{D}[\hat{\sigma}_{l,3+l}]\hat{\rho})$$
$$-\sum_{l=4}^{6} k_{l\to 8}\mathcal{D}[\hat{\sigma}_{8,l}]\hat{\rho} - \sum_{l=1}^{3} k_{7\to l}\mathcal{D}[\hat{\sigma}_{l,7}]\hat{\rho} - k_{8\to 7}\mathcal{D}[\hat{\sigma}_{7,8}]\hat{\rho}$$
$$-\sum_{l=2}^{3}(k_{1\to l}\mathcal{D}[\hat{\sigma}_{l,1}]\hat{\rho} + k_{l\to 1}\mathcal{D}[\hat{\sigma}_{1,l}]\hat{\rho})$$
$$-\sum_{l=5}^{6}(k_{4\to l}\mathcal{D}[\hat{\sigma}_{l,4}]\hat{\rho} + k_{l\to 4}\mathcal{D}[\hat{\sigma}_{4,l}]\hat{\rho})$$
$$-\sum_{l=2}^{3}\frac{1}{2}\chi_{1l}\mathcal{D}[\hat{\sigma}_{l,l} - \hat{\sigma}_{1,1}]\hat{\rho} - \sum_{l=5}^{6}\frac{1}{2}\chi_{4l}\mathcal{D}[\hat{\sigma}_{l,l} - \hat{\sigma}_{4,4}]\hat{\rho}$$
$$-k_{9\to 10}\mathcal{D}[\hat{\sigma}_{10,9}]\hat{\rho} - k_{10\to 9}\mathcal{D}[\hat{\sigma}_{9,10}]\hat{\rho}$$
$$-\sum_{l=4}^{6} k_{l\to 9}\mathcal{D}[\hat{\sigma}_{9,l}]\hat{\rho} - \sum_{l=1}^{3} k_{10\to l}\mathcal{D}[\hat{\sigma}_{l,10}]\hat{\rho} \quad (D3)$$

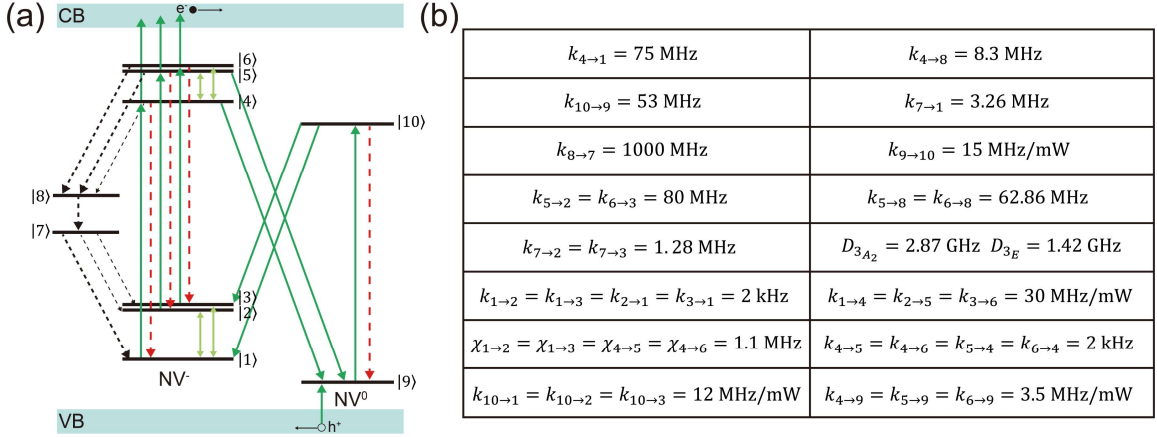

FIG. 8. Labeling of the energy levels of the NV⁻ and NV⁰ (a), and the rates for the various processes (b). For more details, see the text.

Here, the Lindblad superoperator is defined as $\mathcal{D}[\hat{o}]\hat{\rho} = \frac{1}{2}\{\hat{o}^\dagger\hat{o}, \hat{\rho}\} - \hat{o}\hat{\rho}\hat{o}^\dagger$ for any operators $\hat{o}$.

In the following, we describe Eq. (D3) line by line. In the first line, the first and second term in the bracket describes the optical pumping and the spontaneous emission between the triplet states ³A₂, ³E of the NV⁻ center [green and red arrows in Fig. 8(a)] with the rates $k_{l\to 3+l} = 30$ MHz/mW ($l = 1,2,3$), $k_{4\to 1} = 75$ MHz, $k_{5\to 2} = k_{6\to 3} = 80$ MHz. The three terms in the second line describe the inter-system crossing to the singlet state ¹A₁ and from the singlet state ¹E, and the decay between them with the rates $k_{4\to 8} = 8.3$ MHz, $k_{5\to 8} = k_{6\to 8} = 62.8$ MHz, $k_{7\to 1} = 3.26$ MHz, $k_{7\to 2} = k_{7\to 3} = 2.56$ MHz and $k_{8\to 7} = 1$ GHz. In the third and fourth line, the two terms in the first and second bracket describe the spin-lattice relaxation among the spin levels of the triplet state ³A₂ and ³E with the rates $k_{1\to l} = k_{l\to 1} = 2$ kHz ($l = 2,3$), $k_{4\to l} = k_{l\to 4} = 2$ kHz ($l = 5,6$). The two terms in the fifth line describe the spin dephasing among the spin levels of the triplet state ³A₂ and ³E with the rates $\chi_{1l} = 1.1$ MHz ($l = 2,3$) and $\chi_{4l} = 1.1$ MHz ($l = 5,6$). The two terms in the sixth line describe the optical pumping and the spontaneous emission of the NV⁻ and NV⁰ center with the rates $k_{9\to 10} = 15$ MHz/mW, $k_{10\to 9} = 53$ MHz. The two terms in the last line describe the ioniziation of the NV⁻ center and the recombination of the NV⁰ center with the rates $k_{l\to 9} = 3.5$ MHz/mW ($l = 4,5,6$), $k_{10\to l} = 12$ MHz/mW ($l = 1,2,3$).

We solve Eq. (D1) numerically with QuTip package. From the solutions, we can extract the population of various levels $P_l$ ($l = 1, ..., 10$), and then compute the photon emission rate from the NV⁻ and NV⁰ center $F_- = \sum_{l=4}^{6} k_{l\to l-3}P_l$, $F_0 = k_{10\to 9}P_{10}$, from which the total photon emission rate becomes $F = F_- + F_0$. In addition, one can also compute the rate of photon-induced electron and hole carriers $C_e = \sum_{l=4}^{6} k_{l\to 9}P_l$, $C_h = (\sum_{l=1}^{3} k_{10\to l})P_{10}$, from which the total rate becomes $C_t = C_e + C_h$. If one applies an external field through a pair of electrodes, one can obtain a photocurrent proportional to the photon-induce carrier rate $C_t$.

2. Numerical Results

To adapt the above theory for *single NV center* to the experimental system involving *multiple NV*

*centers*, we must account for the influence of the NV center ensemble and the measurement setup.

Due to the presence of multiple NV centers, we need more laser power to excite them than single NV center. To account for this effect, we calculate firstly the fluorescence for single NV center, and fit the curve with the expression $I(P) = I_s \cdot P/(P + P_s)$ to determine the laser power $P_s^t = 0.775$ mW leading to the fluorescence saturation [Fig. 9(a)]. Then, we measure experimentally the fluorescence of the NV ensembles as function of the laser power, and fit it also with the same expression to obtain the saturation laser power $P_s^e = 380$ mW [Fig. 9(b)]. Then, we calculate the ratio $r_{las} = \frac{P_s^t}{P_s^e} \approx 2.04 \times 10^{-3}$, and estimate the optical pumping rate as $k_{l \to 3+l} = k_0 r_{las} P$, where the rate $k_0 = 30$ MHz/mW is for single NV center [48-50] and $P$ is the laser intensity in the experiment.

To estimate the number of involved NV centers, we take the fluorescence image, and fit it with Gaussian function $I(x,y) = A \cdot \exp\left[-\frac{(x-x_0)^2}{2\sigma_x^2}\right] \exp\left[-\frac{(y-y_0)^2}{2\sigma_y^2}\right] + B$, where A denotes the amplitude, $x_0$ and $y_0$ are the center coordinates, $\sigma_x, \sigma_y$ are the standard deviation, B is the background. We fit the fluorescence distribution in two dimension, and obtain an average linewidth of $D \approx 2.55$ μm [Figs. 9(c) and 9(d)]. By assuming a hemispherical shape for the laser spot inside the diamond, we estimate a volume of $V = \frac{1}{2}\frac{4\pi}{3}\left(\frac{D}{2}\right)^3 = 4.34$ μm$^3$. By taking account the measured NV concentration of 3.3 ppb, we estimate the number of NV centers involved as N ≈ 2551.

After determining the relevant parameters, we extend the theoretical model from a single NV center to an NV ensemble. Given the extremely low concentration of NV centers in our diamond [Fig. 7(a)], we assume that there is no interaction between individual NV centers. In Figs. 9(a) and 9(b), we compare the experimentally measured photoluminescence signal with the theoretically calculated one, thereby calibrating the optical pumping rate for the NV ensemble. Furthermore, in Figs. 9(c) and (d), we estimate the laser spot size and estimate the number of NV centers within the illuminated volume. Additionally, the carrier generation rate $R$ is crucial for accurately estimating the magnetic field detection sensitivity. In Fig. 9(e), we show the theoretically calculated the carrier generation rate $R$ for the NV ensemble, which shows a rather good agreement with the one measured with the LIA under laser intensity modulation (stars). In addition, since the conversion from the microwave power to the microwave magnetic field felt by the NV centers depends on the microwave field distribution and the relative position of the NV centers to the antenna, it is rather difficult to estimate and thus is treated as a fitting parameter. By accounting for the above corrections, we have calculated theoretically the PDMR spectra [Fig. 9(f)], and found a good agreement with the experimental data [Fig. 7(a)] under the same conditions. In this way, we reach the good theory-experiment agreement as shown in Fig.4 of the main text.

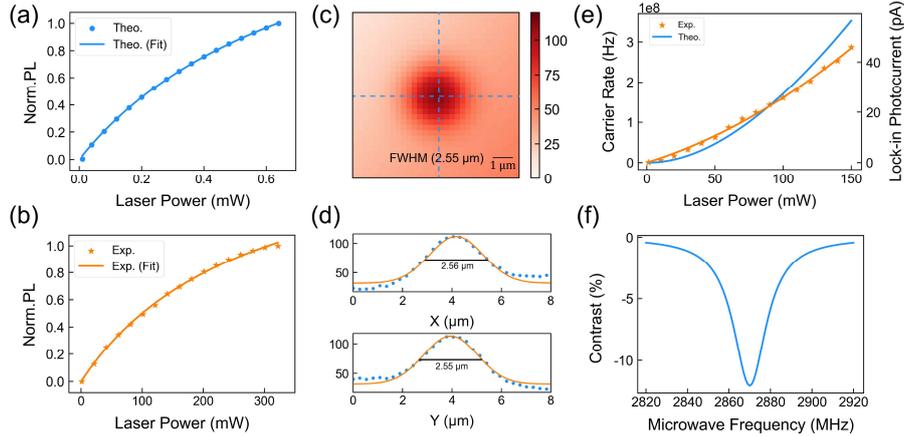

FIG. 9. Determination of key parameters to match the theory to the experimental system. Theoretically calculated photoluminescence (PL, a) and the experimentally measured one (b) as function of the laser power for single NV center and NV ensemble (dots), and the fitting with the expression $I = I_s \cdot P/(P + P_s)$. (c) Fluorescence image of the NV centers taken with a CMOS camera and the fitting of the cut vertical and horizontal lines (d) with Gaussian functions and the extracted linewidths. (e) Theoretically calculated carrier generation rate of NV ensemble, and the experimentally detected value with LIA under laser intensity modulation. (f) Simulated PDMR of NV centers for a laser power 100 mW and a microwave 26 dBm.